\documentclass[smallextended]{svjour3}       
\smartqed
\RequirePackage[OT1]{fontenc}
\RequirePackage[numbers]{natbib}
\RequirePackage[colorlinks,citecolor=blue,urlcolor=blue]{hyperref}
\usepackage{ae,amssymb, amsmath, natbib, graphicx, euscript, mathrsfs, enumerate, empheq, color, lscape, nicefrac, tabularx}
\usepackage{caption}
\usepackage{cmap}
\usepackage{fancyhdr}
\usepackage{pgfplots}
\pgfplotsset{compat=newest}
\usepackage[tikz]{bclogo}
\usepackage{algorithm2e}

\usetikzlibrary{shapes, calc, matrix, positioning, arrows,snakes,backgrounds, patterns, trees}

\newcommand{\uj}{\tilde{u}_j}

\newcommand{\eqd}{\stackrel{Law}{=}}

\renewcommand{\ell}{L}

\newcommand{\E}{{\mathbb E}}

\renewcommand{\i}{\mathrm{i}}

\def\R{\mathbb{R}}
\def\N{\mathbb{N}}

\newcommand{\eps}{\varepsilon}

\newcommand{\argmin}{\operatornamewithlimits{arg\,min}}

\renewcommand{\Re}{\operatorname{Re}}
\renewcommand{\Im}{\operatorname{Im}}

\newcommand{\B}{\mathcal{B}}

\renewcommand{\kappa}{\varkappa}

\newcommand{\I}{{\mathbb I}}

\newcommand{\F}{\mathcal{F}}

\newcommand{\K}{\mathcal{K}}

\usepackage{multirow}
\usepackage{arydshln}

\begin{document}
\title{Modelling the Bitcoin prices and the media attention to Bitcoin via the jump-type processes 
\thanks{ 
}
}

\titlerunning{Bitcoin prices and media attention}

\author{Ekaterina Morozova and Vladimir Panov 
}

\authorrunning{E.~Morozova and V.~Panov} 

\institute{HSE University\\
International Laboratory of Stochastic Analysis and its Applications\\
             Pokrovsky boulevard 11, 109028 Moscow, Russia\newline
             \\                         
	\email{eamorozova@hse.ru \and vpanov@hse.ru
	}
}

\date{Received: \today}
\maketitle

\begin{abstract} In this paper, we present a new bivariate model for the joint description of the Bitcoin prices and the media attention to Bitcoin. Our model is based on the  class of the L{\'e}vy processes and is able to realistically reproduce the jump-type dynamics of the considered time series. We focus on the low-frequency setup, which is for the L{\'e}vy - based  models essentially more difficult than the high-frequency case. We design a semiparametric estimation procedure for the statistical inference on the parameters and the L{\'e}vy measures of the considered processes.  We show that the dynamics of the market attention can be effectively modelled by the L{\'e}vy processes with finite L{\'e}vy measures, and propose a data-driven procedure for the description of the Bitcoin prices. 

\keywords{Bitcoin, media attention, L{\'e}vy process, low-frequency data, deconvolution}
\end{abstract}
\hspace{-0.65cm} \textbf{Mathematics Subject Classiffication} (2020): 60G51; 62M99; 91G30

\setcounter{tocdepth}{3}
\setcounter{secnumdepth}{3}

\section{Introduction}
Stochastic modelling of the Bitcoin prices has been attracted the attention of researchers for the last 10 years. The models, which are able to realistically describe the dynamics of the cryptocurrencies, are quite useful for
many interesting research areas such as the prediction of bubbles, the development of methods to raise individual capital (in particular, portfolio diversification) and many others, see the survey by H{\"a}rdle et al. (\citeyear{Haerdle2020}). 

As usual in the price modelling, the hugest research stream is concentrated on the GARCH-type models and related stochastic volatility models, see, e.g., Dyhrberg (\citeyear{dyhrberg2016}), Katsiampa (\citeyear{katsiampa2017volatility}), Naimy and Hayek (\citeyear{Naimy2018}), Tiwari et al. (\citeyear{Tiwari2019}). It is important to note that  the dynamics of the Bitcoin price essentially differs from the dynamics of many other financial instruments, due to rapid changes (jumps) in the dynamics of returns and volatility. This observation serves as the main motivation of the application of more complicated approaches for modelling the Bitcoin prices.

One of the most promising ideas is to use various characteristics of the media attention to the Bitcoin. In fact, as it was mentioned by many researches, cryptocurrencies are prominently addressed by media, see, e.g.,  Glaser et al. (\citeyear{Glaser14}) and many further references until the end of the article. In this area, the media attention  is quite often measured by  the volume of Google searches (in terms of Search Volume Index, SVI) and Wikipedia requests, see Da et al. (\citeyear{da2011}), Kristoufek (\citeyear{kr2013}), Urquahart (\citeyear{urquhart2018}), Figa-Talamanca and Patacca (\citeyear{figa2019}).

The stochastic factor \(M_t, t \geq 0\), representing the attention index, can be included in the model in different ways.  For instance, in Cretarola et al. (\citeyear{Cretarola}), the process \(M_t\) is described by 
a separate stochastic differential equation, which influences the price process \(S_t, t \geq 0,\) with some shift in time. Namely, the SDE for \(S_t\) includes \(M_{t-\tau}\) with the delay parameter \(\tau,\) which can be estimated by the profile likelihood approach. Nevertheless, this parameter can change over time - in particular,  \(\tau\) can be both positive and negative (changes in media attention can influence the prices of Bitcoin and vice versa). 

In our paper, we introduce a new model for the media attention and the Bitcoin prices. Our approach is based on the class of L{\'e}vy process, which is a well-used tool for the construction of the asset-pricing models, see the brilliant books by Schoutens (\citeyear{Schoutens}), Cont and Tankov (\citeyear{ContTankov}), and the recent papers by Panov and Samarin (\citeyear{panovsamarin}), Gardini et al. (\citeyear{Gardini2021}). 

In the statistical part of this research, we assume that the processes \(M_t\) and \(S_t\) are observed on the equidistant grid \( t=0, \Delta, ..., n \Delta\) with some fixed \(\Delta>0.\) For the L{\'e}vy - based models this case is essentially more difficult than the case of  high - frequency data (i.e., the case \(\Delta \to 0\) as \(n \to \infty\)), see, e.g., the papers by Belomestny and his coauthors (\citeyear{DB}, \citeyear{panov2013d},  \citeyear{BelReiss}, \citeyear{BPW}), Neumann and Reiss (\citeyear{NeuReiss}), Comte and Genon-Catalot (\citeyear{CGC2}), Gugushvili (\citeyear{Gugu}).   Nevertheless, this kind of data allows to avoid the difficulties raised by the  unclear behaviour of the delay parameter (see above), since we consider the increments of prices over some fixed intervals, and it is not important whether the changes in the Bitcoin prices cause the changes in media attention, or vice versa.

\begin{figure}[h]
\includegraphics[width=1\linewidth]{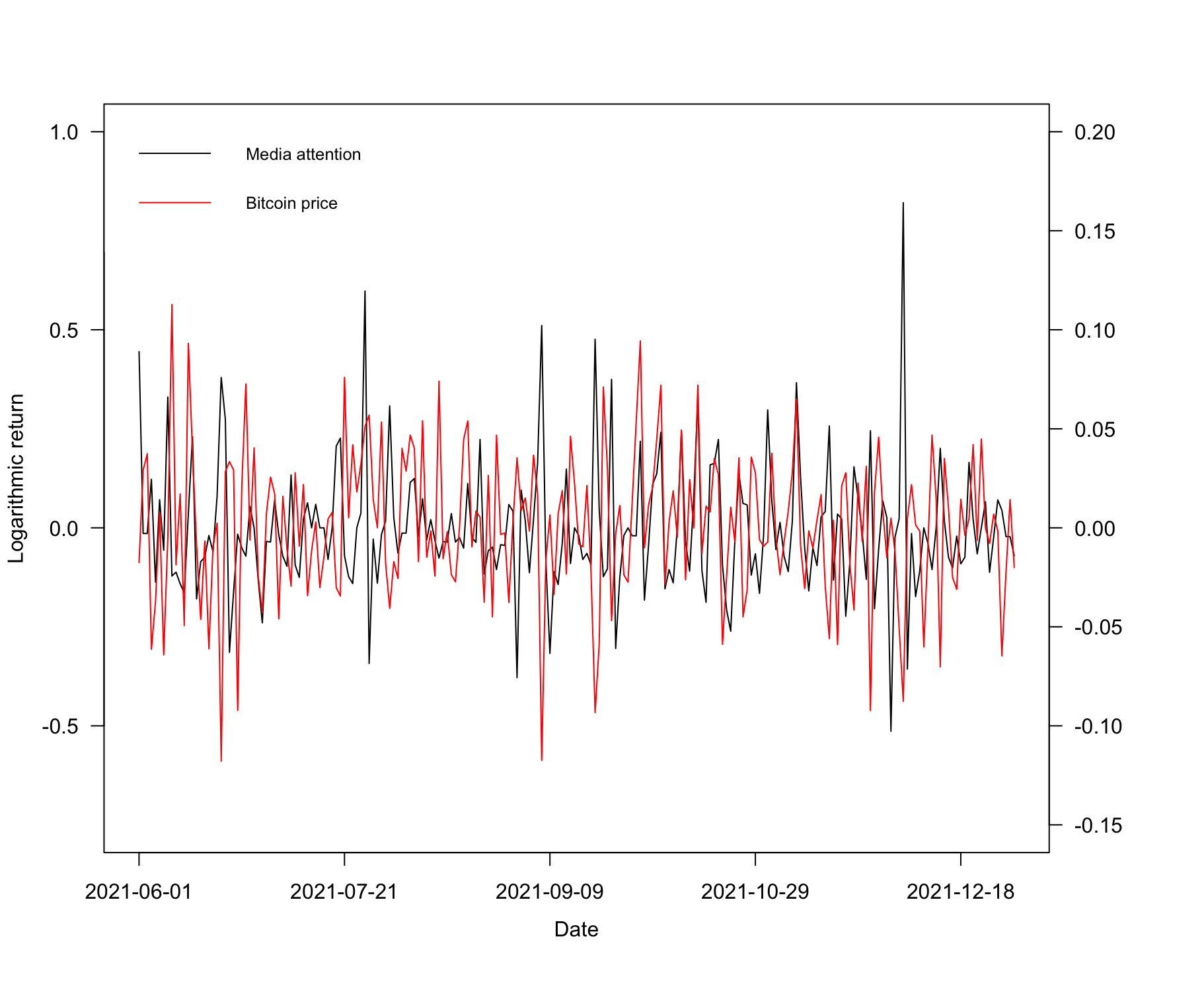}
	\caption{The dynamics of logarithmic returns of media attention and Bitcoin prices in June-December 2021.
	}
	\label{fig:ma_return}
\end{figure}

In what follows, we consider the daily data on logarithmic returns of the Bitcoin price and media attention collected over a five-year period, from 2017 to 2021. Our approach to modelling is based on two key observations. First, let us note that the Bitcoin prices and the attention index to the Bitcoin market (represented by the Google Trends to the word ``Bitcoin") jump on the same intervals of the fixed length \(\Delta=1\) corresponding to the daily data. For example, Figure~\ref{fig:ma_return} represents the dynamics of logarithmic returns of the media attention and the Bitcoin price in six consecutive months of 2021. It can be seen that the spikes in Bitcoin price are well-aligned with those of the media attention: at each time moment, there is typically a rapid change in both values, with the spikes sometimes being headed in opposite directions. This observation is further supported by nonparametric correlation coefficients between the logarithmic returns of media attention and the absolute values for the Bitcoin prices, which are presented in Table~\ref{corr}. As will be shown later, the values of coefficients are slightly higher for the observations that can be classified as jumps.

\begin{table}
\caption{Correlation coefficients for the logarithmic returns of media attention and absolute values of logarithmic returns of the Bitcoin prices.}
\label{corr}
\centering
\begin{tabular}{c|ccc}
& Pearson & Kendall's \(\tau\) & Spearman's \(\rho\) \\ \hline
Coefficient & 0.35357 & 0.173711 & 0.25628 \\ 
p-value & 1.0027\(\cdot 10^{-54}\) & 1.595\(\cdot 10^{-28}\) & 1.0954\(\cdot 10^{-28}\) \\ \hline
\end{tabular}
\end{table}

Another key observation is that the class of  L{\'e}vy processes with finite L{\'e}vy measures can be efficiently used for the description of the media attention, but is no longer appropriate for modelling the Bitcoin prices. Hence, in order to describe the latter, one has to construct another model, taking into account the relationship between the Bitcoin prices and media attention.

These two observations yield the following procedure: 
\begin{enumerate}
\item to build the model for the media attention;
\item to determine the intervals with jumps in the media attention (which coincide with the intervals containing jumps in the Bitcoin prices);
\item to use this information for the construction of the model for the Bitcoin prices.
\end{enumerate}

The paper is organised as follows. In the next section, we focus on the Step~1 and present the jump model for the media attention index.  Subsection~\ref{SI} is devoted to the presentation of the estimation procedure, which is summarised on pages~\pageref{al1} and~\pageref{al2}. The efficiency of the described procedure is illustrated by numerical examples both on the simulated and real data (subsections~\ref{SD1} and \ref{RD1} resp.). Later, in subsection~\ref{SR}, we show that this kind of models is inappropriate for describing the Bitcoin prices. Section~\ref{SMBP} deals with the second step of the aforementioned plan and shows the possibility to determine the jump times. Finally, the model for the  Bitcoin prices is constructed in Section~\ref{SMBP2}.

\section{Stochastic model for media attention}\label{sec_MA}
\subsection{Set-up}
For modelling the attention index to the Bitcoin market (denoted below by \(M_t, t \geq 0\)), we use the class of L{\'e}vy processes with the jump component represented by a compound Poisson process. More precisely,
\begin{eqnarray}\label{model1}
 M_t = \exp\Bigl\{ \mu t + \sigma W_t + \sum_{i=1}^{N_t} \xi_i\Bigr\}, \qquad t \geq 0,
\end{eqnarray}
where \(\mu \in \R, \sigma \in \R_+,\) \(N_t\) is a (homogeneous) Poisson process with intensity \(\lambda\) and \(\xi_1, \xi_2,...\) are i.i.d. random variables with absolutely continuous distribution having the density \(p: \R \to \R_+.\) This class of processes includes the Merton model (corresponding to the case when \(\xi_1\), \(\xi_2\),.. have normal distribution) and the Kou model (double-exponential distribution).

It is known that any L{\'e}vy process  \(L_t, t \geq 0,\) can be represented in the form \(L_t=\mu t + \sigma W_t +J_t\), where \(J_t\) is a pure-jump process, see \cite{Sato}.  As it is shown in many papers (see the references in the introduction), the statistical inference for this general case is rather complicated, provided that the L{\'e}vy measure of \(J_t\) may be infinite and only the low-frequency data are available. Indeed, in this case it is difficult to distinguish the small jumps of  \(J_t\) from the continuous part \(\mu t +\sigma W_t\), since the changes over time interval of the fixed length \(\Delta\) can be caused both by a jump or by the continuous part. Taking into account this fact, we concentrate on  the L{\'e}vy processes with finite L\'evy measure, that is, on the processes in the form~\eqref{model1}.

\subsection{Statistical inference}\label{SI}
Assuming that the 
the process \(M_t\) is observed on the discrete grid \(\{0, \Delta, 2\Delta,...\}\) with fixed \(\Delta>0\) (low-frequency data), we aim to estimate the parameters \((\mu, \sigma, \lambda)\) and the density function \(p(\cdot)\). For the statistical inference we use the method described in \cite{BelReiss}. Note that \[D_k:=\log M_{k\Delta}-\log M_{(k-1)\Delta}, \qquad k=1,2,... \]with \(\log M_0=0,\)  form a sequence of i.i.d. random variables, and due to the L{\'e}vy-Khinchine formula, the characteristic function of their distribution is equal to 
\begin{eqnarray}\nonumber
\phi_\Delta(u) = \E[e^{\i u \log M_\Delta}] 
&=&
\exp\Bigl\{
\Delta\bigl(
\i  \mu u -\frac{1}{2} \sigma^2  u^2 + \int_{\R}\bigl( e^{\i u x} - 1\bigr) \nu (dx) 
\bigr)
\Bigr\}\\ 
&=&
\exp\Bigl\{
\Delta\bigl(
\i \mu u -\frac{1}{2}  \sigma^2 u^2 -\lambda + \lambda\F[p](u) \bigr)\Bigr\}, \qquad u \in \R, \qquad 
\label{LK}
\end{eqnarray}
where \(\nu(B) = \int_B p(u) du, B \in \B(\R),\)  is the L{\'e}vy measure of \(M_t\) and \(\F[p](u) = \int_\R e^{\i u x} p(x) dx\) is the Fourier transform of \(p(\cdot).\) In what follows, we  use the function
\[\varphi_\Delta(u) := \frac{1}{\Delta}\log(\phi_\Delta(u)),  \qquad u \in \R,\]
which can be naturally estimated by 
\begin{eqnarray*} 
\widehat\varphi_\Delta(u) &:=& \frac{1}{\Delta}\log\Bigl(
\frac{1}{n} \sum_{k=1}^n e^{\i u D_k}\Bigr)\\
&=& 
- \frac{1}{\Delta}\log n 
+ 
\frac{1}{2\Delta} \log\Bigl[
\Bigl( \sum_{k=1}^n \cos (u D_k)\Bigr)^2
+
\Bigl( \sum_{k=1}^n \sin ( u D_k)
\Bigr)^2
\Bigr] \\
&&\hspace{4cm}+ \i \frac{1}{\Delta}
 \arctan \left(
\frac{\sum_{k=1}^n \sin ( u D_k)}{\sum_{k=1}^n \cos ( u D_k)}
\right), \qquad u \in \R.\end{eqnarray*}

Due to the Riemann-Lebesque lemma, \(\F[p](u)  \to 0\) as \(|u| \to \infty\). Therefore,  from~\eqref{LK} we get 
\begin{eqnarray*}
\Re\bigl(\varphi_\Delta(u)\bigr) &=& 
 -\frac{1}{2} \sigma^2 u^2  -\lambda + o(1),\\
 \Im \bigl(\varphi_\Delta(u)\bigr)  &=& \mu u +o(1),
\end{eqnarray*}
as \(|u| \to \infty.\)  The first relation yield that for large \(u\) the real part of \(\varphi_\Delta(u)\) is linear in \(u^2,\) and the corresponding coefficients are proportional to \(\sigma^2\) and \(\lambda.\) This observation leads to the following definition of the estimates for \(\sigma^2\) and \(\lambda\):
\begin{eqnarray}
\label{sigmalambda}
\bigl( 
\hat\sigma^2, \hat\lambda
\bigr) 
&=&\argmin_{\sigma^2, \lambda } \int_{\R_+} w^{U_n}(u) 
\Bigl[
\Re\bigl(\widehat\varphi_\Delta(u)\bigr) 
+\frac{1}{2} \sigma^2 u^2  +\lambda 
\Bigr]^2 du \nonumber \\
\label{f1}
&=&\argmin_{\sigma^2, \lambda } \int_{\eps}^1 w(u ) 
\Bigl[
\Re\bigl(\widehat\varphi_\Delta(u U_n)\bigr) 
+\frac{1}{2} \sigma^2 u^2 U_n^2  +\lambda 
\Bigr]^2 du,
\end{eqnarray}
where \(w^{U_n}(u)=U_n^{-1} w(u/U_n)\) with some unbounded increasing sequence of positive numbers \(U_n\) and a function \(w:[\eps,1] \to \R\),  \(\eps>0.\)

Similarly, for the imaginary part of \(\varphi_\Delta(u)\), we also deal with a kind of asymptotic linear regression model. Define the estimate of \(\mu\) as 
\begin{eqnarray}
\label{mu}
\hat\mu
&=& \argmin_{\mu } \int_{\R_+} w^{V_n}(u) 
\Bigl[
\Im\bigl(\widehat\varphi_\Delta(u)\bigr)
-\mu u
\Bigr]^2 du \nonumber \\
\label{f2}
&=& \argmin_{\mu } \int_{\eps}^1 w(u) 
\Bigl[
\Im\bigl(\widehat\varphi_\Delta(u V_n)\bigr)
-\mu u V_n
\Bigr]^2 du
\end{eqnarray}
with some unbounded increasing sequence of positive numbers \(V_n\), which may coincide with \(U_n\). 

Let us summarise the ideas presented above into the following algorithm.

\begin{bclogo}[couleur=blue!15, logo=\bccrayon]
{Algorithm 1: Estimation of \((\mu, \sigma^2, \lambda)\)}
\label{al1}
\begin{algorithm}[H]
\SetAlgoLined

\textbf{Data:}
\(n\) observations of the process \(M_t\) at the equidistant grid \(t=\Delta, 2\Delta, ..., n \Delta.\) 

\vspace{0.3cm}
\textbf{Initiate:} 
Fix  some positive numbers \(U_n, V_{n} \) and \(\eps \in (0,1)\). 
\\
Fix a  function \(w (\cdot) \geq 0\) supported on \([\eps,1]\),\\ 
\hspace{0.2cm}e.g., \(w(x) = \I\{x \in [\eps,1]\}\).  

Fix some \(N \in \N\) and divide \([\eps,1]\) into \(N\) equidistant intervals \(I_j=[u_{j-1}, u_{j}], j=1..N\) with \(u_0 = \eps, u_N=1.\) 

Let \(\tilde{u}_j \in I_j, j=1..N\).

\vspace{0.3cm}
\textbf{Algorithm:}
\begin{enumerate}
\item 
Estimate the function \(\varphi_\Delta(u) := \frac{1}{\Delta}\log(\E[e^{\i u M_\Delta}])\) \\at the points \(u=\uj U_n, j=1..N,\) and \(u=\uj V_n, j=1..N,\) by 
\begin{eqnarray*}
\hat\varphi_\Delta(u) = \frac{1}{\Delta}\log\Bigl(
\frac{1}{n} \sum_{k=1}^n e^{\i u D_k}\Bigr),
\end{eqnarray*}
where \(D_k:=M_{k\Delta}-M_{(k-1)\Delta}, \; k=1..n,\) and \(M_0=0.\)
\item Compute 
\begin{eqnarray*}
\Lambda_{d} &=& \sum_{j=1}^N w(\uj) \uj^{2d}, 
\qquad d=0, 1, 2,\\
\Psi_d &=& \sum_{j=1}^N w(\uj)
\Re\bigl(\widehat\varphi_\Delta(\widetilde{u}_j U_n)\bigr) (\uj U_n)^{2d}, \qquad d=0,1.\\
\Upsilon &=& \sum_{j=1}^N w(\uj)
\Im\bigl(\widehat\varphi_\Delta(\widetilde{u}_j V_n)\bigr) \uj V_n.
\end{eqnarray*}
\item Estimate  \(\sigma^2\) and \(\lambda\) by
\begin{eqnarray*}
\bigl( 
\sigma_n^2, \lambda_n
\bigr) 
&:=&\argmin_{\sigma^2,\lambda} \sum_{j=1}^N w(\widetilde{u}_j) 
\Bigl[
\Re\bigl(\widehat\varphi_\Delta(\widetilde{u}_j U_n)\bigr)
+\frac{1}{2} \sigma^2 \widetilde{u}_j^2  U_n^2 +\lambda 
\Bigr]^2\\
&=&\Bigl(
2 \frac{
\Psi_0 \Lambda_1 U_n^2 
-
\Psi_1\Lambda_0
}{(\Lambda_2  \Lambda_0 - \Lambda_1^2) U_n^4},
\frac{\Psi_1 \Lambda_1 - \Psi_0 \Lambda_2 U_n^2 }
{(\Lambda_2 \Lambda_0 - \Lambda_1^2)U_n^2}
\Bigr)	
\end{eqnarray*}
\item Estimate \(\mu\) by 
\begin{eqnarray*}
\hat{\mu}_n
&:=& \argmin_\mu \sum_{j=1}^N w(\widetilde{u}_j) 
\Bigl[
\Im \bigl(\widehat\varphi_\Delta(\widetilde{u}_j V_n)\bigr)
-\mu \widetilde{u}_j V_n
\Bigr]^2\\ 
&=& \Upsilon / (\Lambda_1 V_n^2).
\end{eqnarray*}
\end{enumerate}
\end{algorithm} 
\end{bclogo}
\begin{remark} \label{rem1}In practice, the choice of \(\eps\) and \(V_n, U_n\)  is based on the idea that for any  \(u \in [\eps U_n, U_n]\) and any \(u \in [\eps V_n, V_n]\), the function \(\varphi_\Delta(u)\) should be well approximated by the following sum 
\begin{eqnarray*}
\varphi_\Delta(u) \approx \i \mu u -\frac{1}{2}  \sigma^2 u^2 -\lambda,
\end{eqnarray*}
and at the same time by its natural empirical estimate 
\begin{eqnarray*}\varphi_\Delta(u) \approx\widehat\varphi_\Delta(u) := \frac{1}{\Delta}\log\Bigl(
\frac{1}{n} \sum_{k=1}^n e^{\i u D_k}\Bigr).
\end{eqnarray*}

\end{remark}
Finally, we turn towards nonparametric estimation of the function \(p.\) From~\eqref{LK}, we get that the natural estimator of \(p\) is 
\begin{eqnarray}
\label{p}
\hat{p}(x) &=& 
\widehat{F}^{-1} \Bigl(\bigl(
\frac{1}{\hat\lambda} 
\bigl(
\widehat\varphi_{\Delta}(u) - \i \hat\mu  u +\frac{1}{2} \hat\sigma^2 u^2
\bigr)
 + 1
 \bigr)
 K\bigl(u / T_n \bigr)
\Bigr) \nonumber \\
&=& \frac{1}{2\pi\hat\lambda} 
\int_{\R}
e^{-\i u x} \bigl(
\widehat\varphi_{\Delta}(u) - \i \hat\mu  u +\frac{1}{2} \hat\sigma^2 u^2
 + \hat\lambda
 \bigr)K\left(u / T_n \right) du,
\end{eqnarray}
where \(K:\R \to \R_+\) is a smoothing kernel defined as \begin{eqnarray}\label{Kx}
K(x) = 
\begin{cases}
1,& |x|\leq 0.05,\\
\exp\left\{-\frac{e^{-1/(|x|-0.05)}}{1-|x|}\right\},& 0.05<|x|<1, \\
0,& |x|\geq 1,
\end{cases}\end{eqnarray}
and
 \(T_n\)  is  an unbounded increasing sequence of positive numbers. 
 
 The particular choice of  the kernel in the form~\eqref{Kx} is not important and any function from a broad class can be chosen. Nevertheless, as it  is shown in~\cite{Belomest2011}, this 
kind of kernels (known as flat-top kernels) is quite useful in semiparametric inference, since it often guarantees the optimality of the  estimators in the minimax sense. Our empirical studies confirm the efficiency of this choice.
\begin{bclogo}[couleur=blue!15, logo=\bccrayon]
{Algorithm 2: Estimation of the L{\'e}vy density \(p:\R \to \R_+\)}
\label{al2}
\begin{algorithm}[H]
\SetAlgoLined

\textbf{Data:}
\(m\) points \(x_1,..., x_m\) (can be chosen by the researcher)

\vspace{0.3cm}
\textbf{Initiate:} 
Fix  a positive numbers \(T_n \). 

\hspace{1.5cm}Divide  \([-1,1]\) into \(N\) equidistant intervals\\\hspace{2cm} \(I_j=[u_{j-1}, u_{j}], j=1..N\) of length \(\delta = 2 / N\)\\ \hspace{2cm} with \(u_0 = -1, u_N=1.\)\\ \hspace{1.5cm} Let \(\tilde{u}_j \in I_j, j=1..N\). 

\vspace{0.3cm}
\textbf{Algorithm:}
\begin{enumerate}
\item 
Estimate the function \(\varphi_\Delta(u) := \frac{1}{\Delta}\log(\E[e^{\i u M_\Delta}])\) \\at the points \(u= \breve{u}_j:=u_j T_n, j=1..N,\) by 
\begin{eqnarray*}
\hat\varphi_\Delta(u) = \frac{1}{\Delta}\log\Bigl(
\frac{1}{n} \sum_{k=1}^n e^{\i u D_k}\Bigr),
\end{eqnarray*}
where \(D_k:=M_{k\Delta}-M_{(k-1)\Delta}, \; k=1..n,\) and \(M_0=0.\)
\item Estimate the density function \(p(x)\) at the points \(x_1,..,x_m\) by 
\begin{eqnarray*}
\label{p}
\hat{p}_n(x_s) &:=& \frac{T_n \delta}{2\pi\hat\lambda_n} 
\sum_{j=1}^N
e^{-\i \breve{u}_j x_s} \bigl(
\widehat\varphi_{\Delta}(\breve{u}_j) - \i \hat\mu_n  \breve{u}_j \bigr. \\
&& \bigl. \hspace{4cm}+\frac{1}{2} \hat\sigma_n^2 \breve{u}_j^2
 + \hat\lambda_n
 \bigr)K\left(\breve{u}_j \right).
\end{eqnarray*}
\end{enumerate}
\end{algorithm} 
\end{bclogo}

\subsection{Simulation study}\label{SD1}

The aim of the current section is to analyse the behaviour of the proposed estimators on the  simulated data. In what follows, we consider the Merton jump-diffusion model, which is defined as~\eqref{model1} with normally distributed jump sizes \(\xi\). For illustrative purposes, we consider the same specification of the model as in~\cite{BelReiss}; namely, the parameters are chosen to be \(\mu=0\), \(\sigma=1\), \(\lambda=10\) and  \(\Delta=0.1\).

\begin{figure}[h]
\includegraphics[width=1\linewidth]{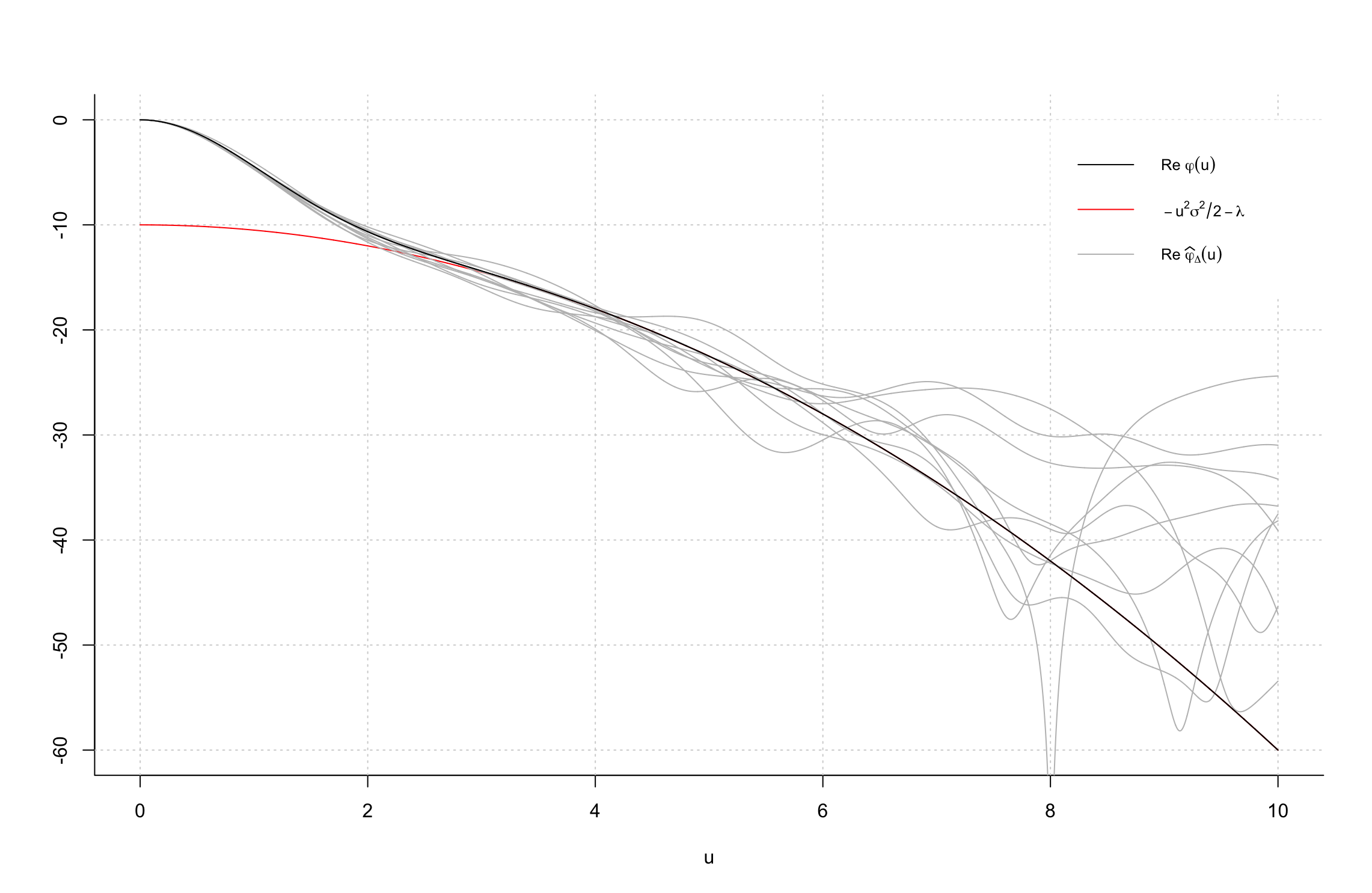}
	\caption{The real part of \(\varphi_{\Delta}(u)\) for the Merton jump-diffusion model, its approximation and the estimates based on 25 repetitions.
	}
	\label{fig:merton_chf}
\end{figure}

Figure~\ref{fig:merton_chf} represents the estimates \(\Re(\widehat{\varphi}_{\Delta}(u))\) based on 25 independent samples of size \(n=1000\) superimposed with the true values of \(\Re(\varphi_{\Delta}(u))\) and its approximation due to the Riemann-Lebesgue lemma. It can be seen that the three lines are fairly close in the interval \(u\in[3,6]\). Hence, as suggested by Remark~\ref{rem1}, the parameters \(\eps\) and \(U_n, V_n\) can be chosen as \(\eps=0.5\) and \(U_n=V_n=6\). 

\begin{figure}[h]
\includegraphics[width=1\linewidth]{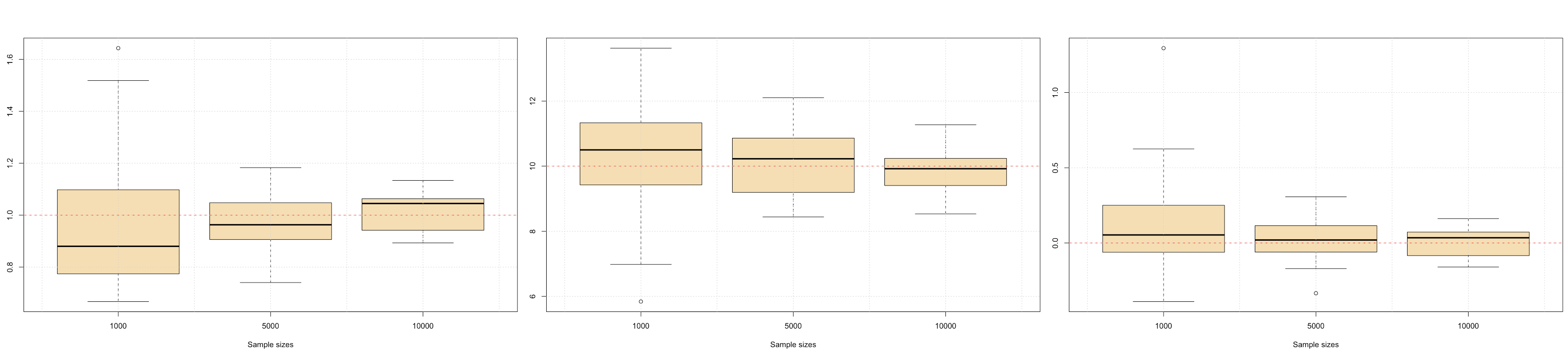}
	\caption{Boxplots for the estimates of \(\sigma^2\) (left), \(\lambda\) (middle) and \(\mu\) (right) for different sample sizes based on 25 repetitions.
	}
	\label{fig:6pars}
\end{figure}

The boxplots for the estimates of the parameters of the model for 25 samples of size \(n=1000, 5000\) and 10000 are presented in Figure~\ref{fig:6pars}. It can be seen that  the estimates get closer to the true value with the growth of \(n\).

\begin{figure}[h]
\includegraphics[width=1\linewidth]{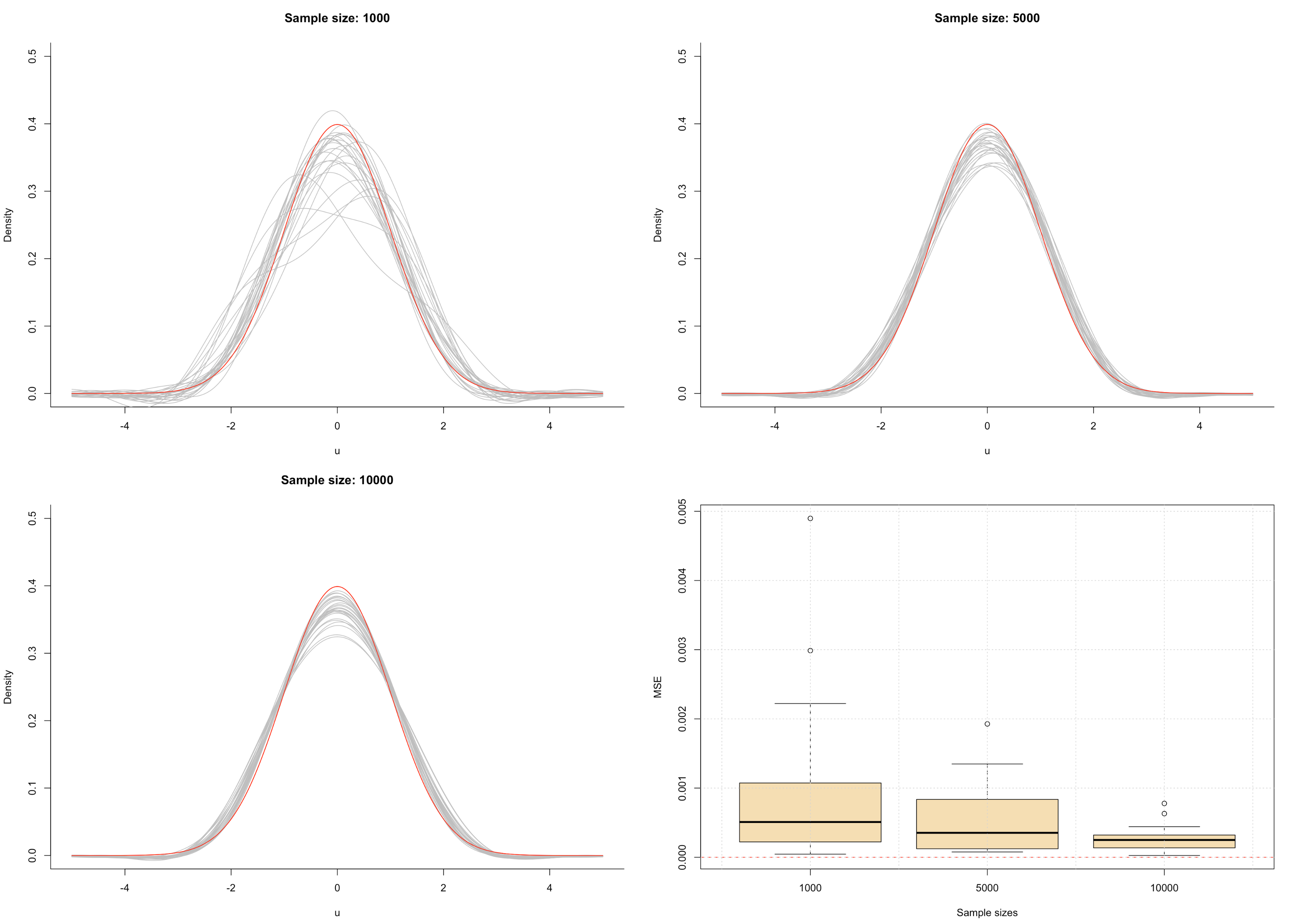}
	\caption{The true density \(p(u) = \frac{1}{\sqrt{2\pi}}e^{-u^2/2}\) of \(\xi\) superimposed with 25 density estimates~\eqref{p} and the corresponding mean-squared errors.
	}
	\label{fig:6densdirect}
\end{figure}

Finally, Figure~\ref{fig:6densdirect} depicts the true density \(p(u) = (2\pi)^{-1/2}e^{-u^2/2}\) of \(\xi\) superimposed with its estimates~\eqref{p} based on 25 independent samples. The last plot represents the boxplots for the values of the mean-squared errors computed as
\[MSE(\hat{p}_n(x))=\frac{1}{m}\sum\limits_{s=1}^{m} \left(\hat{p}_n(x_s)-p(x_s)\right)^2, \]
where the points \(x_m\) are chosen on an equidistant grid from \(-5\) to 5, with \(m=1000\). Again, it can be observed that the estimate becomes more accurate with the growth of \(n\). 

\subsection{Real data}\label{RD1}
The current study considers the daily data over 5 most recent years, from January 1st, 2017 to December 31st, 2021, the total of 1820 observations. The media attention is represented by the Google Trends\footnote{URL:\url{https://trends.google.com/trends}}, which measure the  popularity of a topic relative to the maximum search interest in a given time frame. Namely, the value of 100 corresponds to the most popular topic, while the value of 10 means that it is 10 times less popular. Since the chi-squared test for the logarithmic returns of this characteristic is  resulted into the p-value of 0.2056, the observations within the sample can be assumed to be independent, justifying the use of the models with independent increments and the proposed estimation procedure. 

To begin with, we fit two simple partial cases of the model~\eqref{model1} ---  the Brownian motion with a drift (\(\xi \equiv 0\)), which is purely continuous, and the pure-jump Cauchy model (\(\mu=\sigma=0\) and \(\xi\) having the Cauchy distribution). In both cases the parameters are estimated using the maximum likelihood approach. The Kolmogorov-Smirnov test results into the p-values of \(3.997\cdot 10^{-15}\) and \(8.602\cdot 10^{-6}\) for two models, respectively. As can be seen from Figure~\ref{fig:norm_cauchy_MA}, presented in the appendix, the theoretical distributions indeed do not provide an efficient fit to the true density of logarithmic returns of media attention, since the scale parameter is evidently underestimated in the normal model and overestimated in the Cauchy case. This observation suggests the use of the model~\eqref{model1} in the general setting, which includes both the continuous and the jump components.

\begin{figure}[h]
\includegraphics[width=1\linewidth]{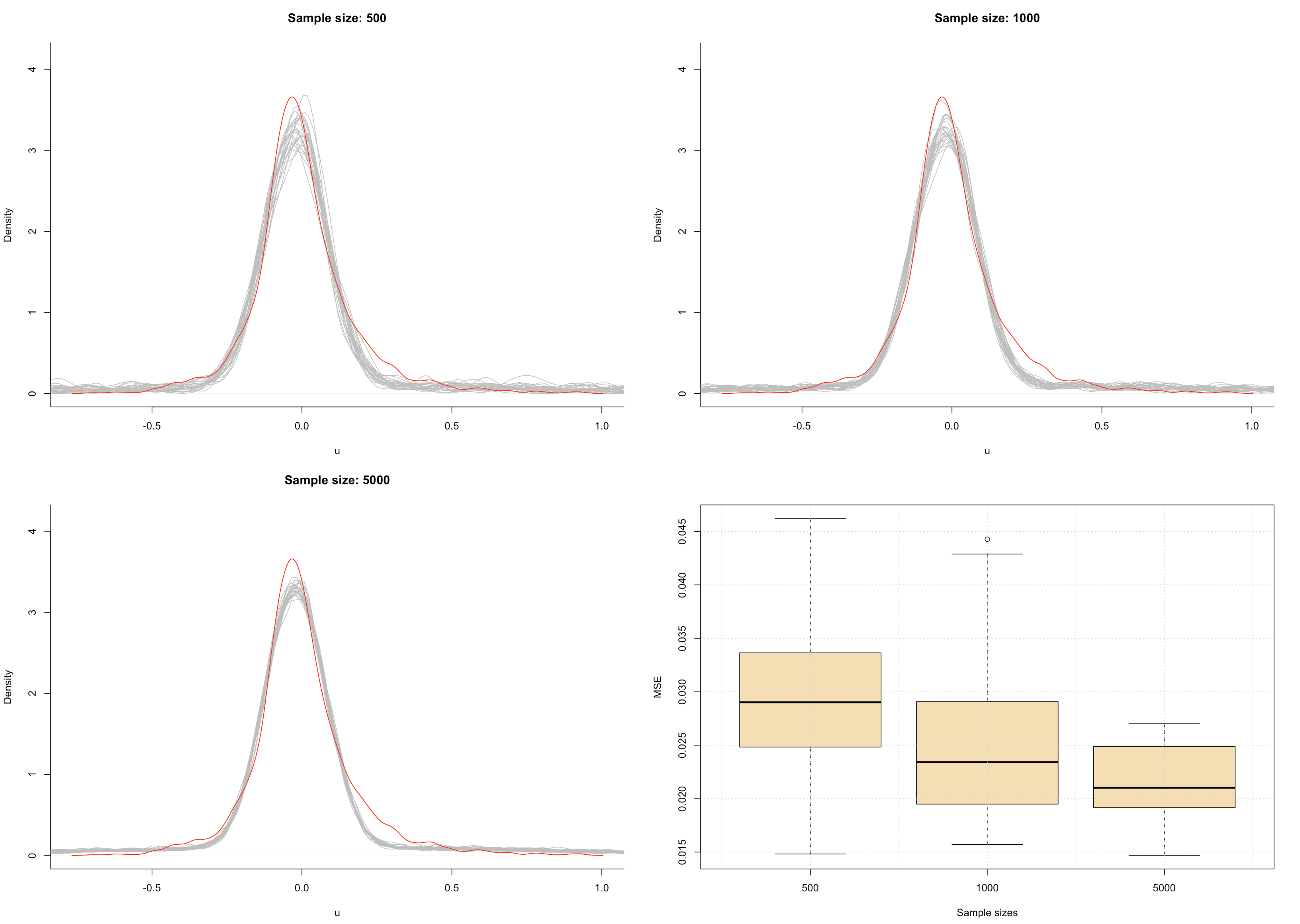}
	\caption{The true density of logarithmic returns of media attention (red) superimposed with densities of 25 simulated samples of different sizes, and the corresponding average quadratic errors.
	}
	\label{fig:fit_MA}
\end{figure}

Now we apply Algorithm~1 for fitting the parameters of the model~\eqref{model1}.  The values of parameters \(U_n\), \(V_n\) and \(\eps\) were chosen as those minimising the average quadratic error of the true and simulated densities based on 25 samples of size 1000. The resulting parameter estimates are \(\hat{\sigma}_n=0.099\), \(\hat{\lambda}_n=0.201\) and \(\hat{\mu}_n=-0.022\). Figure~\ref{fig:fit_MA} represents the true density superimposed with densities of simulated data for different sample sizes (500, 1000 and 5000)  and the optimal values of the parameters equal to  \(U_n=17\), \(V_n=15\) and \(\eps=0.1\). It can be seen that the density curves are very close, and the quadratic errors become smaller as the sample size grows. In addition, the Wilcoxon test does not reject the null hypothesis of zero median between the true and simulated samples with an average p-value of 0.26141 obtained based on 25 simulation runs. Hence, we can conclude that the proposed model indeed provides a reasonable fit to the considered data.

\subsection{Modelling the bitcoin prices: first attempts}\label{SR}
At the end of this section, let us mention that formally the model~\eqref{model1} can also be applied for the description of the Bitcoin prices. Nevertheless, below we provide some empirical evidence that the models of this kind are not appropriate for the considered data. 

As before, we consider the daily data on the same time period, which was obtained from Finam\footnote{URL: \url{https://finam.ru}}. 
Since the chi-squared test resulted into the p-value of 0.1461, the logarithmic returns of the Bitcoin prices were assumed to be independent.

On the first step, the normal and the Cauchy models were fitted to the Bitcoin data. As before, Figure~\ref{fig:norm_cauchy_Bit} (see Appendix) suggests that the first two models do not provide a reasonable fit, with scale parameter being under- and overestimated in the normal and Cauchy models, respectively. The Kolmogorov-Smirnov test also rejects the null of the normal and Cauchy distributions, with p-values of \(8.815\cdot 10^{-14}\) and 0.00341. However, unlike the case of media attention, it turns out that the L\'evy process defined as in~\eqref{model1} is also inappropriate for the Bitcoin price description. Figure~\ref{fig:fit_Bit} depicts the true and simulated densities for values of \(U_n=V_n=44\) and \(\eps=0.5\) chosen as those minimising the discrepancy between the true and simulated densities. It can be observed that even for these parameter values the curves do not seem coherent, and the average quadratic errors both take unreasonably large values and does not decrease with the growth of sample size. Thus, there is an evidence that another  model is needed for the description of the Bitcoin prices.


\section{Determination of the jump times}
\label{SMBP}

As we have already mentioned in the Introduction (see Figure~\ref{fig:ma_return}), the jumps in media attention and the jumps in the Bitcoin prices occur on the time intervals. Let us use the constructed model for the media attention for the determination of the jump times. 

 Assume that on each time interval of length \(\Delta\), only one jump in the media attention may occur. This assumption is partially confirmed to our numerical studies: in fact, since \(\hat\lambda \approx 0.2\) and \(\Delta=1,\) one would expect approximately 1 jump in 5 intervals. Therefore, 
\begin{eqnarray*}
\log M_{k \Delta} - \log M_{(k-1)\Delta} \eqd
\mu \Delta + \sigma \sqrt{\Delta} \zeta + \xi_k J_k, \qquad k=1,...,K
\end{eqnarray*}
where \(\zeta\) is a standard normal random variable and \(J_k\) is an indicator of the jump at time \(k=1,2,...\) From now on, we will think of \(J_k\) as of a non-random value, which is determined by external factors (e.g., external shocks in the market).

\begin{figure}[h]
\includegraphics[width=1\linewidth]{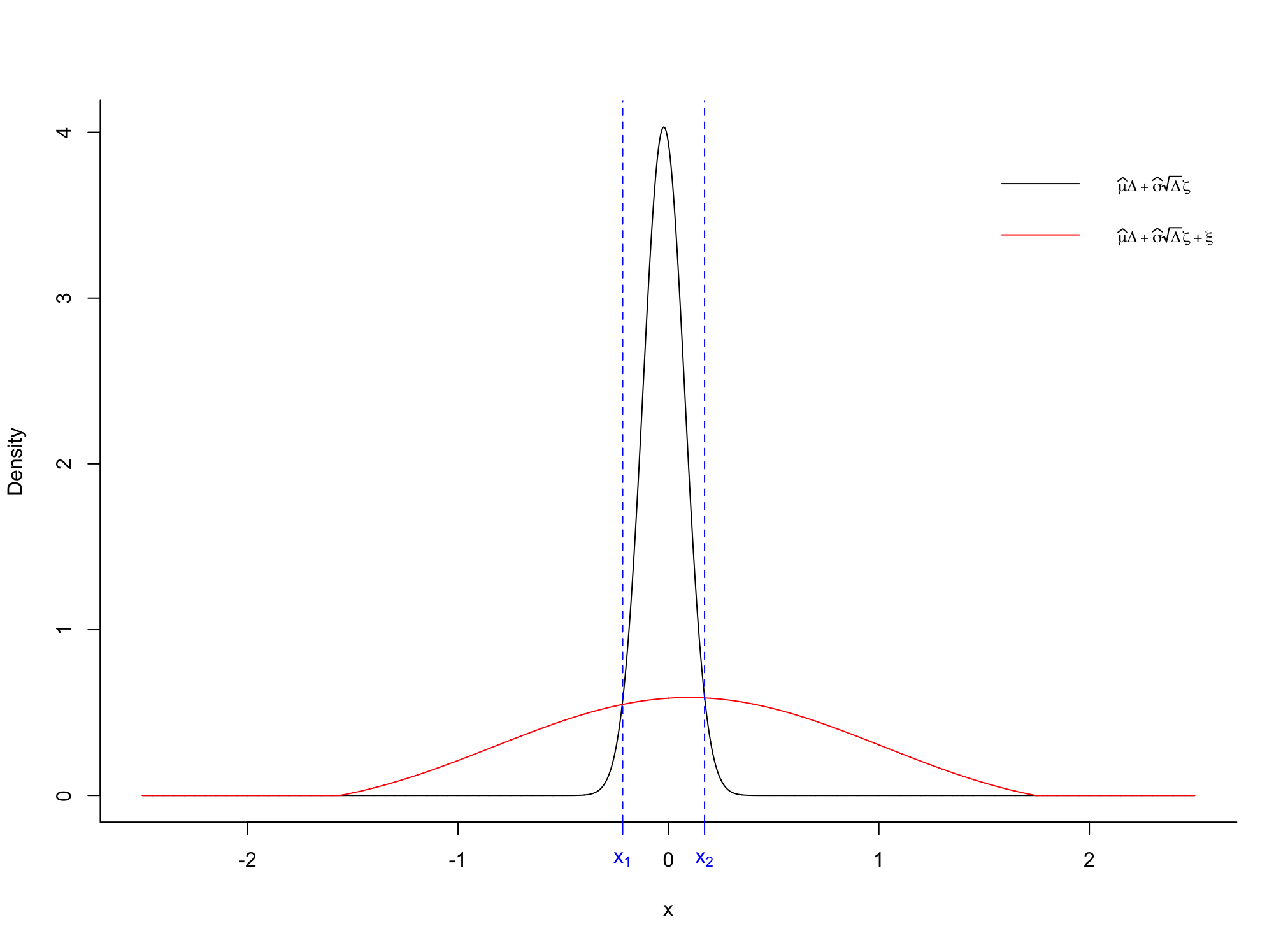}
	\caption{The densities of \(\widehat\mu \Delta + \widehat\sigma \sqrt{\Delta} \zeta\) (black) and of \(\widehat\mu \Delta + \widehat\sigma \sqrt{\Delta} \zeta + \xi\) (red)
	}
	\label{fig:comparison_densities}
\end{figure}

At each time moment \(k=1,2,...\) one should decide between the following 2 possibilities: 
 \begin{eqnarray}
\log M_{k \Delta} - \log M_{(k-1)\Delta} &\eqd&
\widehat\mu_n \Delta + \widehat\sigma_n \sqrt{\Delta} \zeta \label{dec1} \\
\log M_{k \Delta} - \log M_{(k-1)\Delta} &\eqd&
\widehat\mu_n \Delta + \widehat\sigma_n \sqrt{\Delta} \zeta + \xi \label{dec2},
\end{eqnarray}
where the estimates \(\widehat\mu_n\) and \(\widehat\sigma_n\) are defined in Section~\ref{sec_MA},  \(\xi\) is a random variable with a distribution determined by the density function \(\widehat{p}_n,\) \(\xi\) and \(\zeta\) are independent. The densities of the r.v. in the right-hand sides of~\eqref{dec1} and~\eqref{dec2} are depicted in Figure~\ref {fig:comparison_densities}.

This figure suggests the following data-driven decision rule. All values of media attention between \(x_1 \approx -0.2175\) and \(x_2 \approx 0.1715\) are assumed to belong to the first distribution, and for these values we set \(J_k = 0\). To the contrary, observations that are smaller than \(x_1\) or greater than \(x_2\), which correspond to the atypically large or small values for the model~\eqref{dec1}, are assumed to come from the distribution with the jump component, and the value of \(J_k\) is set equal to 1. Hence, the set \[\mathcal{K}= \Bigl\{ k = 1..K:  \qquad J_k=1\Bigr\},\] corresponds to the jump times, while its complement \(\K^c= \{1..K\} \backslash \K\) represents the time moments without jumps.

\begin{figure}[h]
\includegraphics[width=1\linewidth]{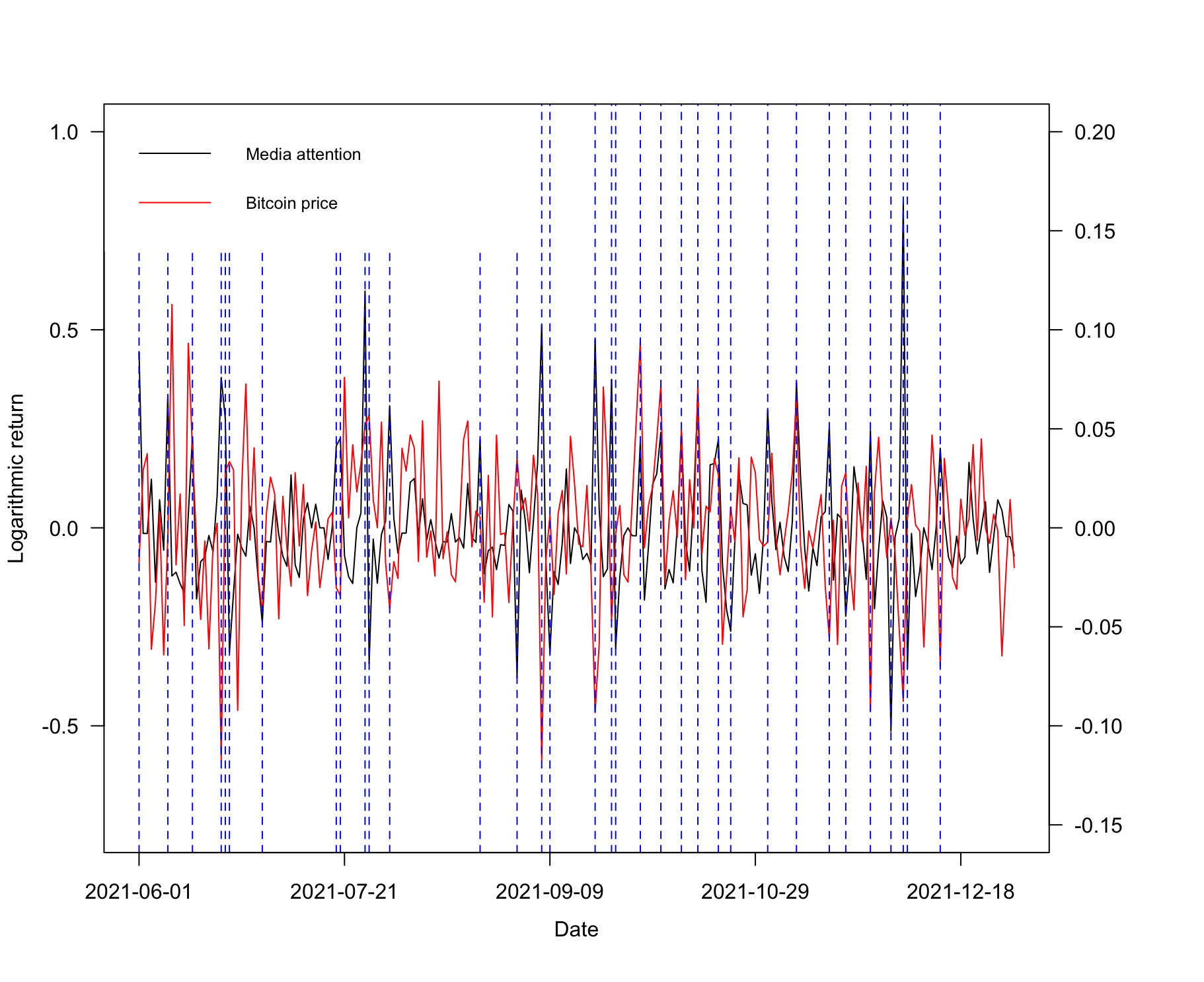}
	\caption{The jumps in logarithmic returns of media attention and Bitcoin prices in June-December 2021
	}
	\label{fig:ma_return_jumps}
\end{figure}

Figure~\ref{fig:ma_return_jumps} represents the jump times from June till December 2021 determined by the procedure described above. It can be observed that the estimated jump times indeed correspond to the significant changes both in media attention and the Bitcoin prices. The total number of jump times constituted 337, which is coherent with the results of Section~\ref{sec_MA}, demonstrating that the jumps should occur approximately once in four days. The values of correlation coefficients computed for the observations classified as jumps are presented in Table~\ref{corr2}. It can be noticed that the values of these coefficients, all being significant, are greater than those presented in Table~\ref{corr} and computed for all observations in total. Hence, there is indeed a significant dependence between the jumps of two processes.

\begin{table}
\caption{Correlation coefficients for the logarithmic returns of media attention and absolute values of logarithmic returns of the Bitcoin price, computed only for the time moments classified as jumps.}
\label{corr2}
\centering
\begin{tabular}{c|ccc}
& Pearson & Kendall's \(\tau\) & Spearman's \(\rho\) \\ \hline
Coefficient & 0.44357 & 0.3026 & 0.43869 \\ 
p-value & 1.1204\(\cdot 10^{-17}\) & 1.2176\(\cdot 10^{-16}\) & 2.77314\(\cdot 10^{-17}\) \\ \hline
\end{tabular}
\end{table}


\section{Modelling the Bitcoin prices}\label{SMBP2}

In view of the results of the previous two sections, we consider the following model for the Bitcoin prices \(S_{k\Delta}, \; k=1..K,\)
\begin{eqnarray*}
\log\bigl( 
S_{k \Delta} 
\bigr)
-
\log\bigl( 
S_{(k-1) \Delta} 
\bigr)
=\widetilde{\mu} \Delta + \widetilde{\sigma}\bigl(\widetilde{W}_{k \Delta}
-
\widetilde{W}_{(k-1) \Delta}
\bigr) + \eta_k \I \bigl\{ k \in \K \bigr\}, \,
\end{eqnarray*}
where \(\widetilde{\mu} \in \R,  \widetilde{\sigma} \in \R_+\) are two constant parameters, \(\widetilde{W}\) is a Brownian motion, and \(\eta_1, \eta_2,...\) are i.i.d. random variables independent on \(\widetilde{W}\). The set \(\mathcal{K}\) is determined above in Section~\ref{SMBP}.
So, the jumps in the Bitcoin price occur at the same times as those of the media attention.

\subsection{Estimation of the continuous part} 
Due to our procedure, the sample is divided into two parts, corresponding to the indices from the set \(\mathcal{K}\) and  to its compliment~\(\K^c\). In what follows, we will use the second part of the sample for the estimation of the parameters \(\widetilde{\mu}\) and \(\widetilde{\sigma}\), and the first part --- for the estimation of the density of \(\eta_1.\) 

Since the sample 
\[\log\bigl( 
S_{k \Delta} 
\bigr)
-
\log\bigl( 
S_{(k-1) \Delta} 
\bigr), \qquad  k \in \K^c,\] 
is normally distributed with the mean \(\widetilde{\mu}\Delta\) and the variance \(\widetilde{\sigma}^2 \Delta,\) we estimate the parameters \(\widetilde{\mu}\) and \(\widetilde{\sigma}\) 
by the maximum likelihood approach, which leads to \(\widehat{\widetilde{\sigma}}=0.037\) and \(\widehat{\widetilde{\mu}}=0.002\). Table~\ref{CIcont} presents the real values of the first three moments, standard deviation and median for the continuous part of the Bitcoin returns along with the confidence intervals obtained for the normal distribution. It can be observed that the confidence intervals have a small length and cover the true values. The null hypothesis of the Wilcoxon test is not rejected with a p-value of 0.32404, obtained as an average over 25 simulated samples of size 1000 as before.

\begin{table}
\setlength{\tabcolsep}{4.7pt}
\caption{99\% confidence intervals for the moments, standard deviation and median of the continuous part of the logarithmic returns of Bitcoin prices.}
\label{CIcont}
\centering
\begin{tabular}{c|ccccc}
& 1st moment\(\cdot 10^{3}\) & 2d moment\(\cdot 10^{3}\) & 3d moment\(\cdot 10^{3}\) & SD\(\cdot 10^{3}\) & Median\(\cdot 10^{3}\) \\ \hline
Lower CI & \(-0.5017\) & 1.3515 & \(-0.0027\) & 35.2067 & \(-1.0035\)\\
True value & 1.9634 & 1.3612 & 0.0104 & 36.854 & 2.0861 \\
Upper CI & 4.4285 & 1.3708 & 0.0234 & 38.7042 & 5.1756 \\ \hline
\end{tabular}
\end{table}





\subsection{Estimation of the jump part}
\begin{figure}[h]
\includegraphics[width=1\linewidth]{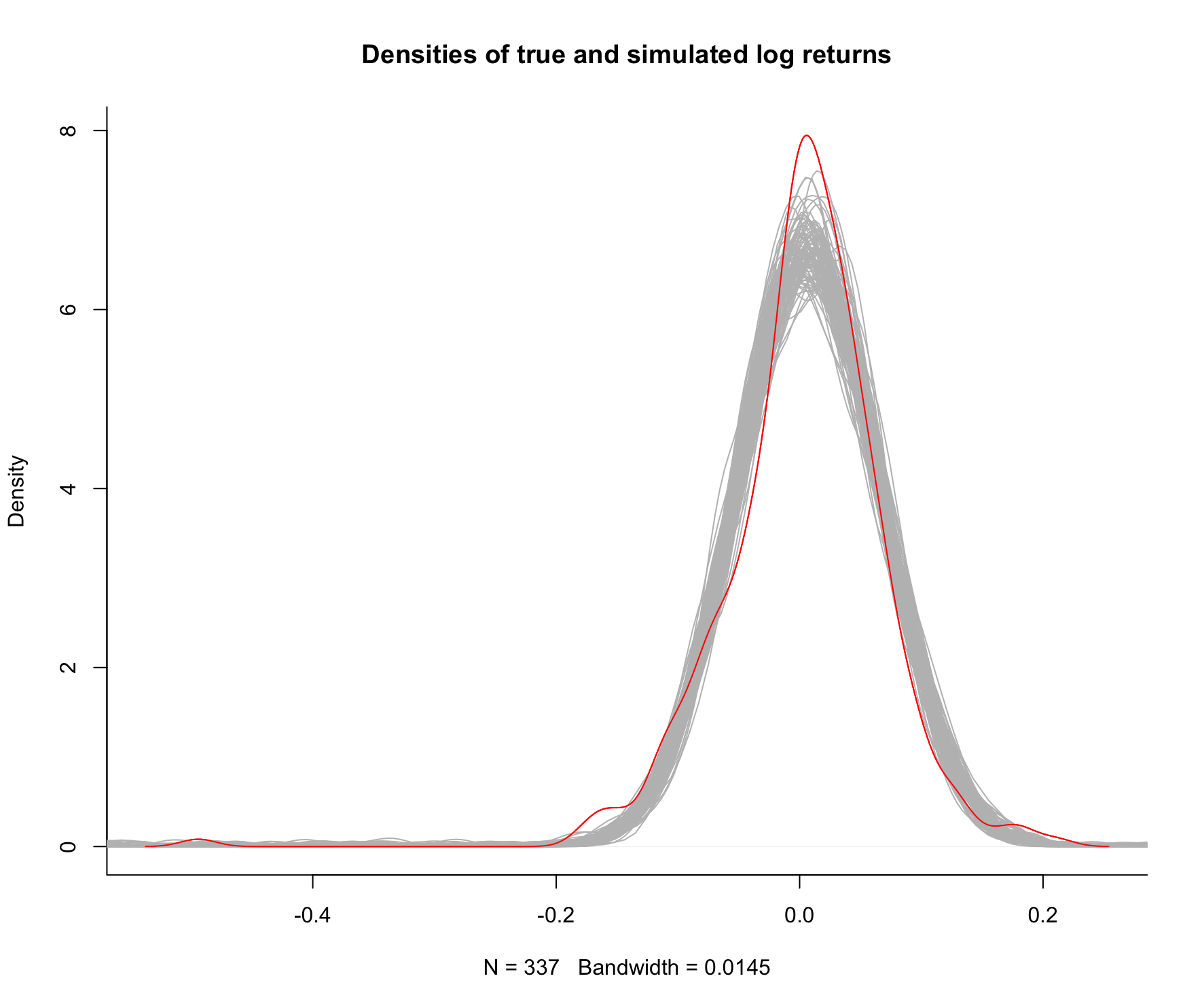}
	\caption{The true density of jumps in logarithmic returns of the Bitcoin price superimposed with densities of 25 simulated samples of size 1000.
	}
	\label{fig:jumppart_dens}
\end{figure}

Similar to the case of media attention, we assume that the jump part of the Bitcoin price can be described by a L\'evy process. That is, for each \(k\in\K\) the characteristic function \(\phi_{\eta_k - \eta_{k-1}} (u)\) admits a representation
\[\phi_{\eta_k - \eta_{k-1}} (u) = \exp\left\{\Delta\left(\i\mu^{\circ} u-\frac{1}{2}(\sigma^{\circ})^2u^2-\lambda^{\circ} + \lambda^{\circ}\F[p^{\circ}](u)\right)\right\},\]
where as before, \(\sigma^{\circ}>0, \mu^{\circ}\in\R\) are some parameters, and \(p^{\circ}\) is the density of pure jumps. Thus, we can apply the estimation procedure described in Section~\ref{SI}. As before, the values of the technical parameters \(U_n\), \(V_n\) and \(\eps\), defined as in Section~\ref{SI}, are determined as those minimising the mean-squared error between the real and estimated densities, and turn out to be \(U_n=15\), \(V_n=23\) and \(\eps=0.6\). The parameter estimates are \(\sigma^{\circ}=0.0576\), \(\lambda^{\circ} =0.01362\) and \(\mu^{\circ}=0.0071\). Figure~\ref{fig:jumppart_dens} represents the true density of jumps in logarithmic returns of the Bitcoin price and simulated densities obtained from 25 independent samples of size 1000. It can be seen that the densities are very close. The Wilcoxon test does not reject the null hypothesis with an average p-value of 0.611. In addition, Table~\ref{CIjump} indicates that the confidence intervals for moments, standard deviation and median include the true parameter values. Hence, we conclude that the proposed model provides a good fit to the jump part of the Bitcoin price. 

\begin{table}
\setlength{\tabcolsep}{4.7pt}
\caption{99\% confidence intervals for the moments, standard deviation and median of the jump part of the logarithmic returns of Bitcoin price}
\label{CIjump}
\centering
\begin{tabular}{c|ccccc}
& 1st moment & 2d moment & 3d moment & SD & Median \\ \hline
Lower CI & \(-0.0074\) & 0.0044 & \(-0.0991\) & 0.0658 & \(-0.0001\)\\
True value & 0.0028 & 0.0044 & \(-0.0003\) & 0.0662 & 0.0062 \\
Upper CI & 0.0119 & 0.0381 & 0.0442 & 0.1951 & 0.0111 \\ \hline
\end{tabular}
\end{table}

\section{Discussion}
The results presented in this paper follow the promising and rather popular in the literature idea to use the indicators of the media attention for the prediction --- or at least description --- of the Bitcoin prices. We consider the  stochastic models with independent increments, namely, the L\'evy processes.
Using the recent data, we provide some empirical evidence that the L{\'e}vy processes with finite L{\'e}vy measure (e.g., as in the Kou and the Merton models) describe well the daily returns of media attention measured by Google Trends, but are insufficient for modelling the returns of the Bitcoin price. Hence, we develop a model for the latter indicator using the observation that the jumps in media attention and the Bitcoin price are closely linked. Determining the jump times from the model constructed for media attention, we describe the Bitcoin price by modelling the jump and the continuous parts separately.  As shown by various statistical tests, the proposed approach provides a good fit to the considered data, allowing to successfully capture its various statistical characteristics.

\appendix
\section*{Appendix}

\begin{figure}[h]
\includegraphics[width=1\linewidth]{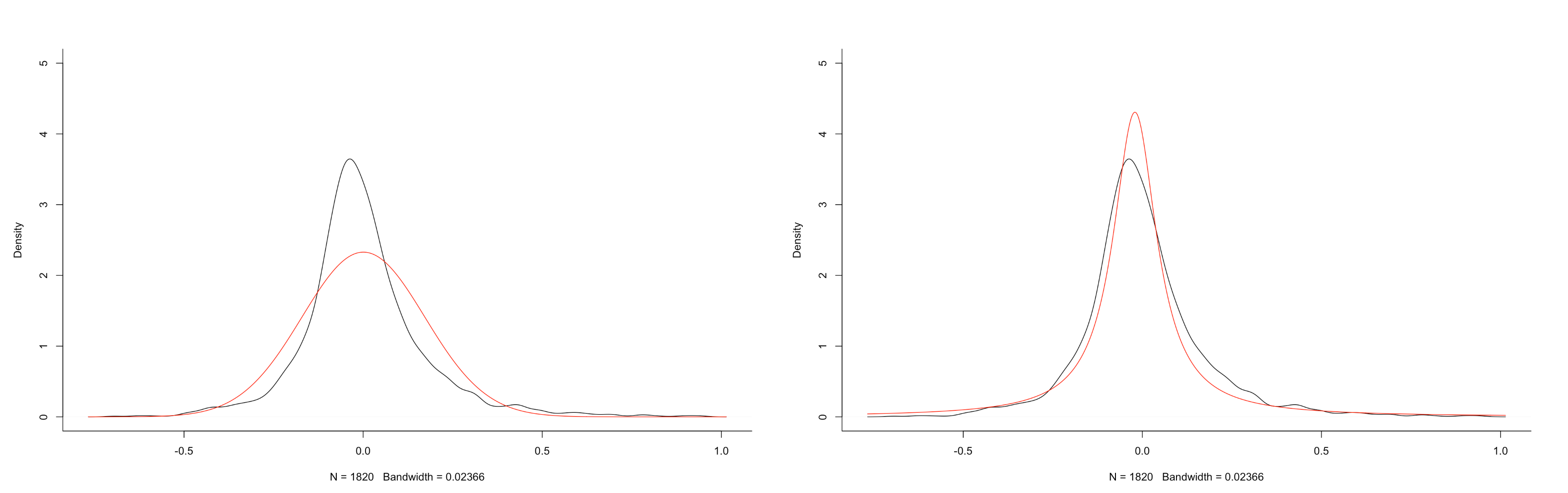}
	\caption{The true density of logarithmic returns of media attention (black) superimposed with the normal density (left) and Cauchy density (right) with estimated parameters
	}
	\label{fig:norm_cauchy_MA}
\end{figure}

\begin{figure}[h]
\includegraphics[width=1\linewidth]{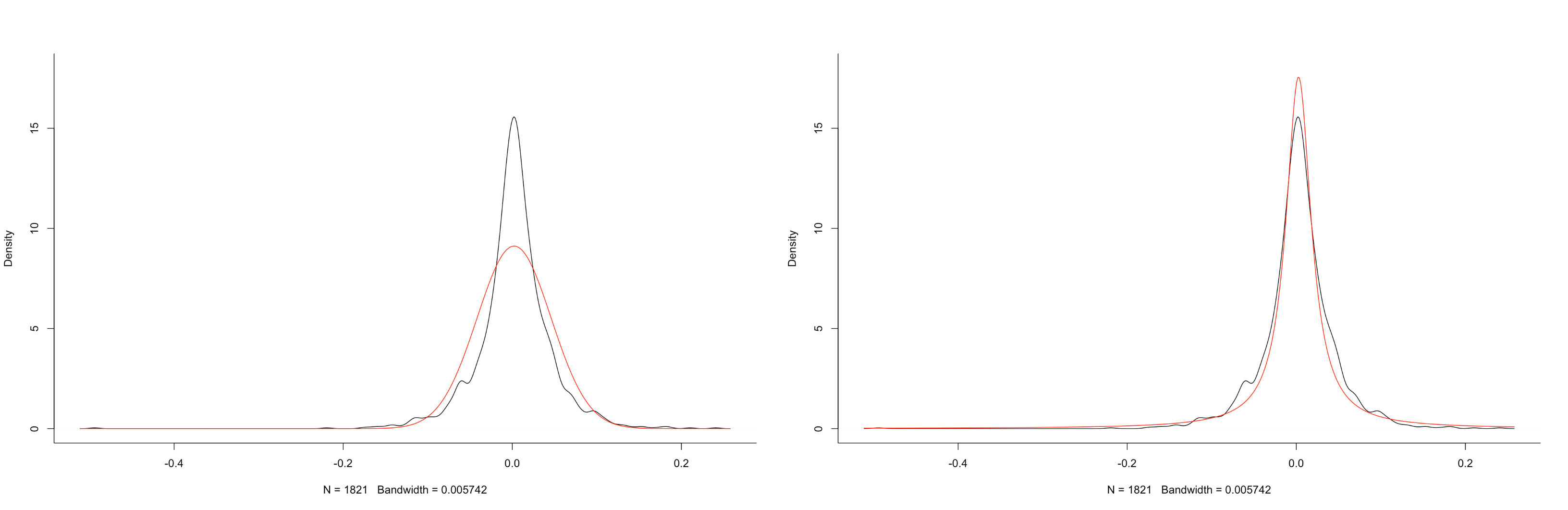}
	\caption{The true density of logarithmic returns of Bitcoin price (black) superimposed with the normal density (left) and Cauchy density (right) with estimated parameters
	}
	\label{fig:norm_cauchy_Bit}
\end{figure}

\begin{figure}[h]
\includegraphics[width=1\linewidth]{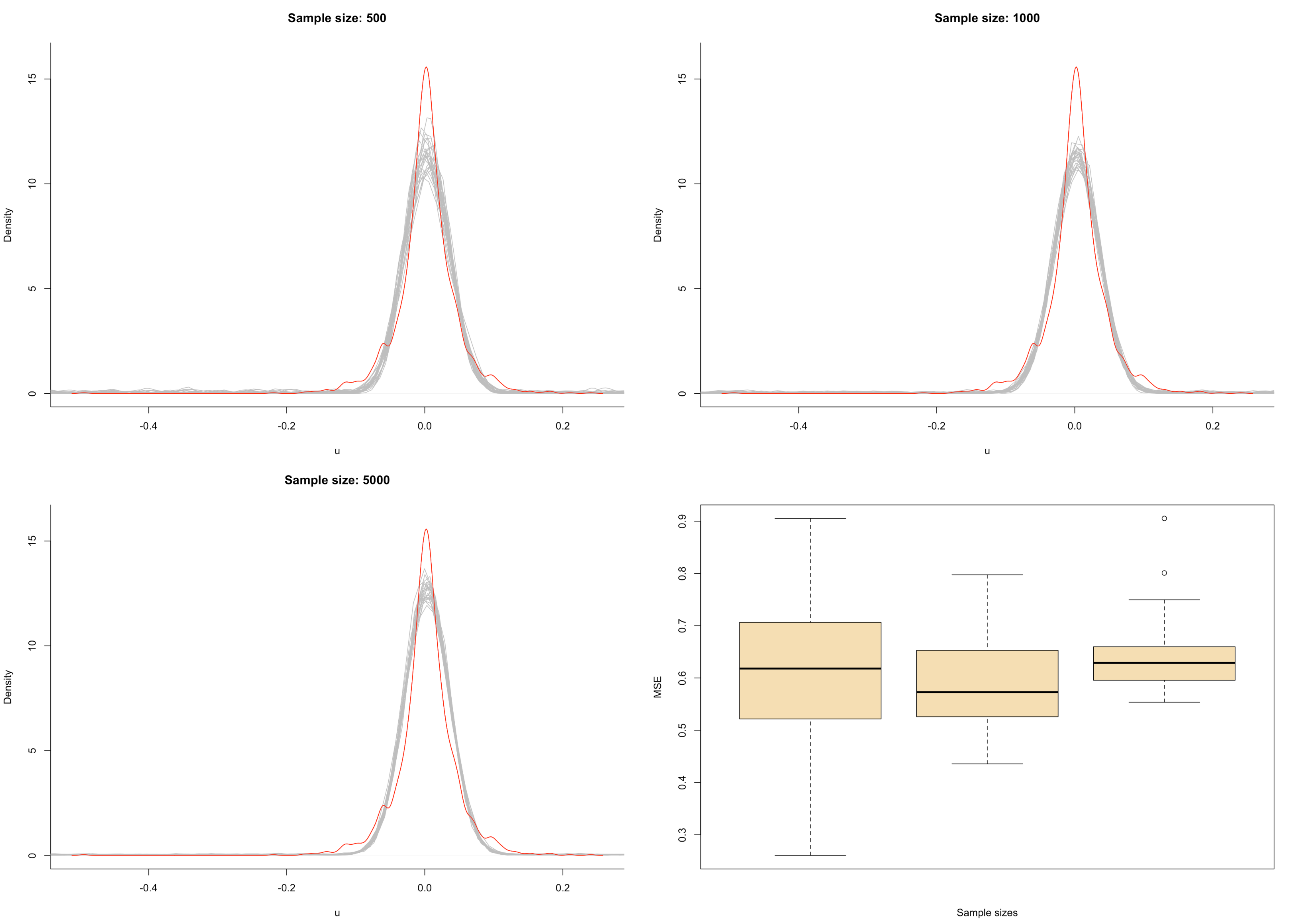}
	\caption{The true density of logarithmic returns of Bitcoin price (red) superimposed with densities of 25 simulated samples of different size, and the corresponding average quadratic errors
	}
	\label{fig:fit_Bit}
\end{figure}

\bibliographystyle{spbasic}
\bibliography{Panov_bibliography}
\end{document}